# Machine Learning for a Sustainable Energy Future

Zhenpeng Yao[1,2,8,9†*], Yanwei Lum[4,5†], Andrew Johnston[5†], Luis Martin Mejia-Mendoza[3], Xin Zhou[6], Yonggang Wen[6], Alán Aspuru-Guzik[3,7,8]*, Edward H. Sargent[5]*, and Zhi Wei Seh[4]*

[1]Center of Hydrogen Science, Shanghai Jiao Tong University, 800 Dongchuan Road, Shanghai 200240, China
[2]Chemical Physics Theory Group, Department of Chemistry and Department of Computer Science, University of Toronto, Toronto, Ontario M5S 3H6, Canada
[3]Institute of Materials Research and Engineering, Agency for Science, Technology and Research (A*STAR), 2 Fusionopolis way, Innovis, Singapore 138634, Singapore
[4]Department of Electrical and Computer Engineering, University of Toronto, Toronto, Ontario, M5S 3G4, Canada
[5]School of Computer Science and Engineering, Nanyang Technological University, 50 Nanyang Avenue, Singapore 639798, Singapore
[6]Vector Institute for Artificial Intelligence, Toronto, Ontario M5S 1M1, Canada
[7]Canadian Institute for Advanced Research (CIFAR) Lebovic Fellow, Toronto, Ontario M5S 1M1, Canada

[8]Innovation Center for Future Materials, Zhangjiang Institute for Advanced Study, Shanghai Jiao Tong University, 429 Zhangheng Road, Shanghai 201203, China
[9]The State Key Laboratory of Metal Matrix Composites, School of Materials Science and Engineering, Shanghai Jiao Tong University, 800 Dongchuan Road, Shanghai 200240, China

[†]These authors contributed equally to this work.
[*]Correspondence: z.yao@sjtu.edu.cn, alan@aspuru.com, ted.sargent@utoronto.ca, and sehzw@imre.a-star.edu.sg.

## Abstract

Transitioning from fossil fuels to renewable energy sources is a critical global challenge; it demands advances – at the levels of materials, devices, and systems – for the efficient harvesting, storage, conversion, and management of renewable energy. Researchers globally have begun incorporating machine learning (ML) techniques with the aim of accelerating these advances. ML technologies leverage statistical trends in data to build models for prediction of material properties, generation of candidate structures, optimization of processes, among other uses; as a result, they can be incorporated into discovery and development pipelines to accelerate progress. Here we review recent advances in ML-driven energy research, outline current and future challenges, and describe what is required moving forward to best lever ML techniques. To start, we give an overview of key ML concepts. We then introduce a set of key performance indicators to help compare the benefits of different ML-accelerated workflows for energy research. We discuss and evaluate the latest advances in applying ML to the development of energy harvesting (photovoltaics), storage (batteries), conversion (electrocatalysis), and management (smart grids). Finally, we offer an outlook of potential research areas in the energy field that stand to further benefit from the application of ML.

## Introduction

Combustion of fossil fuels, used to fulfill ~80% of the world's energy needs, is the largest single source of rising greenhouse gas (GHG) emissions and global temperature.[1] Increased utilization of renewable sources of energy, notably solar and wind power, is an economically viable path to meet the climate goals of the Paris Agreement.[2] However, the rate at which renewable energy has grown has been outpaced by ever-growing energy demand, and as a result the fraction of total energy produced by renewable sources has remained constant since 2000.[3] It is thus essential to accelerate the transition towards sustainable sources of energy.[4] Achieving this requires energy technologies, infrastructure, and policies that enable and promote the harvest, storage, conversion, and management of renewable energy.

In sustainable energy research, suitable candidates (*e.g.*, photovoltaic materials) must first be chosen from the combinatorial space of possible materials, followed by synthesis at a high enough yield and quality for use in devices (*e.g.*, solar panels). The timeframe of a representative materials discovery process is 15-20 years[5,6] – leaving considerable room for improvement. Furthermore, the devices have to be optimized for robustness and reproducibility to be incorporated into energy systems (*e.g.*, solar farms),[7] where management of energy usage and generation patterns is needed to further guarantee commercial success.

Here we explore the extent to which machine learning (ML) techniques can help to address many of these challenges.[8–10] ML models can be used to predict specific properties of new materials without the need for costly characterization; they can generate new material structures with desired properties; they can understand patterns in renewable energy usage and generation; and they can help inform energy policy by optimizing energy management at both device and grid levels.

In this review, we will give an overview of the key concepts of ML driven energy materials discovery, management, and deployment. We then introduce Acc(X)eleration Performance Indicators (XPIs), which can be used to measure the effectiveness of accelerated energy materials platforms. Next, we discuss closed-loop ML frameworks and evaluate the latest advances in applying ML to the development of energy harvesting, storage, and conversion techniques, as well as the integration of ML into a smart power grid. Finally, we offer an outlook of critical research directions in the field that stand to further benefit from ML.

## Machine learning glossary: essential concepts

With the availability of large datasets[11,12] and boosted computing power, various ML algorithms have been developed to solve diverse problems in energy. Below, we provide a brief overview of the types of problems ML can solve in energy technology, and then summarize the status of ML driven energy research. We include a glossary of important ML terms (**Tab. S1**). More detailed information of the nuts and bolts of the ML techniques can be found in previous reviews.[13–16]

### *Property prediction*

Supervised learning models are predictive (or discriminative) models which are given a datapoint *x,* and seek to predict a property *y* (*e.g.*, band gap[17]) after being trained on a labelled dataset. The property *y* can be either continuous or discrete. They have been used to aid or even replace physical simulations or measurements under certain circumstances.[18,19]

### *Generative materials design*

Unsupervised learning models are generative models which can generate or output new examples *x*' (*e.g.*, new molecules[20]) after being trained on an unlabelled dataset. This can be further enhanced with additional information (physical properties) to condition or bias the generative process: this allows the models to generate examples with improved properties, leading to the property-to-structure approach called inverse design.[21,22]

### *Self-driving labs*

Self-driving or autonomous labs[23] use ML models to plan and perform experiments, including the automation of retrosynthesis analysis (*e.g.*, reinforcement learning aided synthesis planning[24,25] ), prediction of reaction products (*e.g.*, convolutional neural networks for reaction prediction[26,27]), reaction condition optimization (*e.g.*, active learning optimized robotic workflow[23,28–33]). Self-driving labs are a key component to the "closed-loop" inverse design.[21]

*Empowering characterization*

ML models have been used to aid the quantitative or qualitative analysis of experimental observations and measurements, including crystal structure determination from transmission electron microscopy (TEM) images,[34] identifying coordination environment[35] and structural transition[36] from X-ray absorption spectroscopy and crystal symmetry in electron diffraction[18].

*Accelerating theoretical computations*

ML models can enable otherwise intractable simulations: ML can reduce the computational cost for systems with increased length and time scales,[37,38] and provide a lack of feasible potentials and functionals for complex interactions, *etc*.[39]

*Optimizing system management*

ML models can aid the management of energy systems at a device or grid power level by predicting lifetimes (*e.g.*, battery life[40,41]), adapting to new loads (*e.g.*, long-short term memory for building load prediction[42]), and optimizing performance (*e.g.*, reinforcement learning for smart grid control[43]).

## Acc(X)eleration Performance Indicators (XPIs)

Many recent reports discuss ML-accelerated approaches to discovering materials and managing energy systems. As a result of this research effort, we posit that there should be a consistent baseline from which these reports can be compared. For energy systems management, performance indicators at the device, plant and grid levels have been reported,[44,45] yet there are no equivalent counterparts for accelerated materials discovery.

The primary goal in materials discovery is to develop efficient materials that are ready for commercialization. The commercialization of a new material takes intensive research efforts that can span up to two decades: the goal of every accelerated approach should be to accomplish this in an order of magnitude less time. The materials science field can benefit from studying an analogous case of vaccine development. New vaccines historically take 10 years[46] from conception to market. However, in less than one year from the start of the COVID-19 pandemic, several companies were able to develop and begin releasing vaccines. This achievement was in part due to an unprecedented global research intensity, but also by a shift in the technology: DNA-sequencing underwent a paradigm shift in 2008, and the cost of sequencing DNA began decreasing exponentially, significantly faster than Moore's Law,[47,48] enabling researchers to screen orders-of-magnitude more vaccines than was previously possible.

ML for energy technologies has many commonalities with other fields like biomedicine. They both are the extensive application scenarios of ML advances from the computer algorithms field, sharing the same methodology and principles. However, the difference does exist when talking about the practice of employing ML in different fields. Attacking distinct problems may expose the model to extra unique requirements. For example, ML models for medical applications have to build a complex structure to enable regulatory oversight to ensure the safe development, use, and monitoring of systems, which usually won't happen to the energy field.[1] Meanwhile, data availability varies significantly from field to field that biomedical researchers can work with a relatively large amount of well-accumulated data which energy people usually lack. And the limited accessibility to a sufficiently large amount of data could constrain the usage of sophisticated ML models that can have more capabilities (such as deep learning models). However, adaption has been rather quick among all fields with a rapidly increased number of groups recognizing the importance of statistical methods and starting to use them for various problems. We posit that the use of high-throughput experimentation (HTE) and ML in materials discovery workflows can result in a similar paradigm shift but will first need a set of metrics by which they can be evaluated and compared, so that they may better improve.

Accelerated materials discovery methods should be judged on the time it takes for a new material to be commercialized. We recognize that this is not a useful metric for new platforms, nor is it one that can be used to quickly decide which platform is best suited for a particular scenario. To this point, we propose here Acc(X)eleration Performance Indicators (XPIs) that new materials discovery platforms should report (**Fig. 1**).

*Acceleration factor (AF) of new materials*

This XPI will be evaluated by dividing the number of new materials that are synthesized and characterized per unit time with the accelerated platform by the number of materials that are synthesized and characterized with traditional methods. For example, an AF of 10 means that for a given time period, the accelerated platform can evaluate 10x more materials than a traditional platform. For materials with multiple target properties, researchers should report the rate-limiting AF.

*Number of new materials with threshold performance*

This XPI tracks the number of new materials discovered with an accelerated platform that have a performance greater than baseline value. The selection of this baseline value is critical – it should be something that fairly captures the standard to which new materials need to be compared. As an example, an accelerated platform that seeks to discover new perovskite solar cell materials should track the number of devices made with new materials that have a better performance than the current record solar cell.[49]

*Performance of best material over time*

This XPI tracks the absolute performance – whether it is Faradaic efficiency, power conversion efficiency, or other – of the best material as a function of time. For the accelerated framework, this should follow a trajectory which grows more rapidly than the counterpart of the traditional methods.[50]

*Repeatability and reproducibility of new materials*

This XPI seeks to ensure that the new materials discovered are consistent and repeatable – this is a key consideration for commercialization and can be used to screen out materials that otherwise would only fail at the commercialization stage. The performance of a new material should not vary by more than $x$% of its mean value: if it does, this material should not be included in either XPI-2 or XPI-3.

*Human cost of the accelerated platform*

This XPI reports the total costs of the accelerated platform. This should include the total number of researcher hours that were needed to: design and order the components for the accelerated system; develop the programming and robotic infrastructure; develop and maintain databases used in the system; and maintain and run the accelerated platform. This metric will provide researchers with a realistic estimate of the resources required to adapt an accelerated platform for their own research.

Each of these XPIs can be measured for computational, experimental, or integrated accelerated systems. Consistently reporting each of these XPIs as new accelerated platforms are developed will better allow researchers to evaluate the growth of these platforms and can provide a consistent metric by which different platforms can be compared. As a demonstration, we applied the XPIs to evaluate the acceleration performance of several typical platforms: Edisonian-like trial-test, Robotic photocatalysis development[23], and DNA encoded library based kinase inhibitor design[51] **(Tab. S2)**. As the reference, the Edisonian-like approach has a calculated overall XPIs score around 1, while the most advanced method among them, the DNA encoded library-based drug design, exhibits a score/acceleration factor of $10^7$. For the sustainability field, the Robotic photocatalysis platform shows an overall XPIs score of $10^5$, catching up with the biological counterpart.

## Closed-loop machine learning frameworks for materials discovery

In this subsection, we talk about the status and the future of the materials design workflow **(Fig. 2)**. The traditional approach to materials discovery is often Edisonian-like, relying on trial and error to develop materials with specific properties. Firstly, a target application is identified, and a starting pool of possible candidates is selected **(Fig. 2a)**. The materials are then synthesized and incorporated into a device or system to measure their properties. These results are then used to establish empirical structure and property relationships, which guide the next round of synthesis and testing. This slow process goes through as many iterations as required and each cycle could take several years to complete.

A computation-driven, high-throughput screening approach **(Fig. 2b)** offers a faster turnaround. To explore the overall vast chemical space ($\sim 10^{60}$ possibilities), human intuition and expertise can be used to create a library with enlarged number of materials systems of interest (*e.g.*, $\sim 10^4$). Theoretical calculations are carried out on these candidates and the top performers (*e.g.*, $\sim 10^2$) are then experimentally verified. With luck, the material with the desired functionality is 'discovered'. Otherwise, this process is repeated in another region of the chemical space. However, this approach can still be very time-consuming and computationally expensive and the region of the chemical space sampled is small.

ML presents an attractive option for significantly increasing the chemical space sampled, without costing extra time and effort. ML is data-driven, screening datasets with the objective of detecting patterns, which are the physical laws that govern the system of interest. In this case, these laws correspond to materials structure and property relationships. This workflow involves high-throughput virtual screening **(Fig. 2c)** and begins by selecting a larger region (*e.g.*, $\sim 10^6$) of chemical space of possibilities *via* human intuition and expertise. Theoretical calculations are carried out on a representative subset (*e.g.*, $\sim 10^4$) and the results are used for training a discriminative ML model. The model can then be used to make predictions on the other candidates in the overall selected chemical space ($\sim 10^6$).[10] Similarly, the top $\sim 10^2$ candidates are experimentally verified, and the results are used to improve the predictive capabilities of the model in an iterative loop. If the desired material is not 'discovered', the process is repeated on another region of chemical space.

An improvement on the previous approaches is a framework that requires limited human intuition or expertise to direct the chemical space search: the automated virtual screening approach **(Fig. 2d)**. Firstly, the region of chemical space can be picked at random to initiate the process. Thereafter, this process is similar to the previous approach, except that the computational and experimental data is also used to train a generative ML model. This generative model solves the 'inverse' problem: given a required materials property, the goal is to predict an ideal structure and composition in the chemical space. This enables a directed, automated search of the chemical space, towards the final goal of 'discovering' the ideal material.[9]

## Machine learning for energy harvesting, storage, and conversion

In this subsection, we will explore how ML has so far been used to accelerate the development of materials and devices for energy harvesting (photovoltaics), storage (batteries), and conversion (electrocatalysis). Besides all the examples discussed here, we also prepared two collection tables that summarize the grand challenges in sustainable materials research (**Tab. 1**) and the details of key studies (**Tab. 2**).

## Accelerating design of photovoltaics

Here we outline the rapid progress that has been made by deploying ML to accelerate the discovery of new optoelectronic materials and devices for photovoltaics, and outline the major challenges associated with each step (**Fig. 3, 4**).

*Photovoltaics materials discovery*

One materials class for which ML has proved particularly effective is perovskites, as these materials have a vast chemical space from which the constituents may be chosen. Early representations of perovskite materials for ML were atomic-feature representations, in which each structure is encoded as a fixed-length vector comprised of an average of certain atomic properties of the atoms in the crystal structure.[52,53] A technique similar to this was used to predict new lead-free perovskite materials with the proper bandgap for solar cells.[54] These representations allowed for high accuracy but did not account for any spatial relation between atoms.[55,56] Materials systems can also be represented as images[57] or as graphs,[17] which can treat systems with diverse sizes. The latter representation is particularly compelling, as perovskites, particularly organic-inorganic perovskites, have crystal structures that incorporate a varying number of atoms, as the organic molecules can vary in size.

While bandgap prediction is an important first step, it is not sufficient to only use these parameters as an indicator of a useful optoelectronic material. Other parameters – including electronic defect density and stability – are equally important. Defect energies are addressable with computational methods, but the calculation of defects in structures is extremely computationally expensive which inhibits the generation of a dataset of defect energies from which an ML model can be trained. Broberg *et al.* developed a library to expedite the high-throughput calculation of defect energies.[58] This library will be pivotal in developing a database of defect energies in semiconductors. Researchers can then use ML to predict both the formation energy of defects *and* the energy levels of these defects. This will ensure that the materials selected from a high-throughput screening will not only have the correct bandgap but will also either be defect tolerant or defect resistant and can find use in commercial optoelectronic devices.

Even without access to a large dataset of experiments, ML can accelerate the discovery of optoelectronic materials. Using a self-driving laboratory approach,[59] the authors demonstrate that the number of experiments required to optimize an organic solar cell can be reduced from 500 experiments to just 60. The robotic synthesis method accelerates the rate of learning of the ML models and also drastically reduces the cost of the chemicals needed to run the optimization.

### *Solar device structure and fabrication*

Photovoltaic devices require optimization of layers other than the active layer to maximize performance. One component is the top transparent conductive layer, which needs to have both high optical transparency as well as high electronic conductivity.[60,61] Sun and co-workers employed a genetic algorithm to optimise the topology of a light-trapping structure, which enabled a broadband adsorption efficiency of 48.1%, which is more than a 3-fold increase over the Yablonovitch limit.[62]

ML was also used to develop better models to determine optimal band gaps for solar cell operation,[63] where a single standard spectrum is usually used by researchers. Actual solar irradiance fluctuates based on factors such as the position of the sun, atmospheric phenomena and the season. Ripalda *et al.* proposed a solution to this issue by using ML to reduce yearly spectral sets into a few characteristic spectra,[63] allowing for the calculation of optimal bandgaps for real-world conditions.

Demant and co-workers used a CNN to predict the I-V characteristics of as-cut Si wafers based on their photoluminescence (PL) images[64]. An artificial neural network was used by Honysz and co-workers to predict the contact resistance of metallic front contacts for silicon solar cells, which is critical for the manufacturing process[65].

## Enhancing electrochemical energy storage

Electrochemical energy storage is an essential component in electric vehicles (EVs), consumer electronics, stationary power stations, and so on. State-of-art electrochemical energy storage features several solutions that have varying efficacy in different applications: lithium-ion batteries (LIBs) exhibit excellent energy density and are widely used in electronics and EVs; redox flow batteries (RFBs) have drawn significant attention for use in stationary power storage. ML approaches have been widely employed in the field of batteries, including for new materials discovery (*e.g.*, solid-state ion conductors[66–68], redox active electrolyte for RFBs[69]), battery management (*e.g.*, state of charge determination,[70] state of health,[71,72] remaining life prediction[40,41]). However, there exist opportunities for further breakthroughs enabled by ML (**Fig. 3, 4**).

### *Electrode and electrolyte materials design*

Developing new Li-ion battery materials that can deliver higher operating voltages, energy densities and longer lifetimes is of paramount interest. Currently, layered oxide materials (*e.g.,* $LiCoO_2$, $LiNi_xMn_yCo_{1-x-y}O_2$) have been used extensively as cathode materials for alkali metal-ion (Li/Na/K) batteries; and yet universal design principles for new battery materials remain undefined. Joshi *et al.* used data from the Materials Project to predict electrode voltages for different materials in alkali metal-ion batteries (Na and K)[73]. They used their model to generate voltage profile diagrams and proposed 5000 different electrode materials with appropriate voltages. Sendek *et al.* applied ML for discovery of solid Li-ion conducting materials[74,75]. They were able to screen 12,000 candidates and discovered 10 new Li-ion conducting materials.

Flow batteries consist of active materials dissolved in electrolytes that flow into a cell with electrodes that facilitate redox reactions, and organic flow batteries are of particular interest. In organic flow batteries, the solubility of the active material in the electrolyte and the charge/discharge stability dictate performance. ML methods have explored the chemical space to find suitable electrolytes for organic redox flow batteries[76,77]. S. Kim *et al.* proposed a multi-kernel-Ridge regression method to accelerate the discovery of organic molecules using multiple feature training[76]. The authors predicted the solubility dependence on pH in anthraquinone with different numbers and combinations of sulfonic and hydroxyl groups. Future opportunities lie in the exploration of large combinatorial spaces for inverse design of high entropy electrodes[78] and high voltage electrolytes.[79] To this end, Sanchez-Lengeling[21] proposed deep generative models to assist the discovery of new materials based on SMILES representation of molecules.

*Battery device and stack management*

A combination of mechanistic and semi-empirical models is currently used to estimate capacity and power loss in lithium-ion batteries; however, the models are only applicable to specific failure mechanisms or situations and cannot predict the lifetimes of batteries at an early stage. Severson *et al.* present a solution to this by developing a mechanism-agnostic model based on ML that can predict accurately battery cycle life, even at the early stage of a battery's life.[40] A combined early-prediction and BO model was used to rapidly identify the optimal charging protocol with the longest cycle life.[41] ML can be used to accelerate the optimization of batteries for longer lifetimes; it remains to be seen if these models can be generalized to a different battery chemistries.

While ML is able to predict the lifetime of batteries, it is difficult to identify the underlying degradation mechanisms and correlate them to the state of health to gain meaningful insights. To this end, incorporation of domain knowledge into a hybrid physics-based ML models can improve explainability and reduce overfitting.[80] However, it remains challenging to incorporate the physics of battery degradation processes into a hybrid model: representation of electrode materials that encode both compositional and structural information is far from trivial. Validation of these models also require the development of *operando* characterization techniques, such as liquid-phase TEM, that reflect true operating conditions as closely as possible.[81]

ML methods can also predict important properties of battery storage facilities. Kandasamy *et al.* used a neural network to predict the charge/discharge profile in two types of stationary battery systems, lithium iron phosphate and vanadium redox flow batteries[82]. Battery power management techniques must also consider the uncertainty and variability that arise from both the environment and the application. An iterative Q-learning (reinforcement learning) method was also designed for battery management and control in smart residential environments[83]. Given the residential load and the real-time electricity rate, the method is effective at optimizing battery charging/discharging/idle cycles. Murphey *et al.*[84] proposed discriminative neural network based models to optimize battery usage in electric vehicles.

## Targeted discovery of electrocatalysts

Electrocatalysis enables the conversion of simple feedstocks (*e.g.,* water, carbon dioxide and nitrogen) into valuable chemicals/fuels (*e.g.,* hydrogen, hydrocarbons and ammonia), using renewable energy as an input.[85] The reverse reactions are also possible in a fuel cell: hydrogen can be consumed to produce electricity.[86] Active and selective electrocatalysts must be developed to improve the efficiency of these reactions.[87,88] Here, we summarize how ML has been used to accelerate electrocatalyst development and device optimization (**Fig. 3, 4**).

*Electrocatalyst materials discovery*

The most common descriptor of activity is the adsorption energy of intermediates on a catalyst[88,89]. While these adsorption energies can be calculated using DFT, catalysts possess multiple surface binding sites, each having different adsorption energies[90]: the number of possible sites increases dramatically if alloys are considered and thus becomes intractable under conventional means[91].

DFT calculations are critical to the search for electrocatalytic materials[92] and efforts have been made to accelerate this process and reduce the computational cost by using surrogate ML models.[38,39,93,94] Complex reaction mechanisms, involving hundreds of possible species and intermediates can also be simplified using ML, with a surrogate model to predict the most important reaction steps and deduce the most likely reaction pathways[37]. Recent work has also shown how ML can be used to screen for active sites across a random, disordered nanoparticle surface[95,96]. DFT calculations were performed on only a few representative sites, which were then used to train a neural network to predict the adsorption energies of all active sites.

Catalyst development can benefit from high-throughput systems for catalyst synthesis and performance evaluation.[97,98] An automatic ML driven framework was developed to screen a large intermetallic chemical space for $CO_2$ reduction and $H_2$ evolution.[99] The model predicts the adsorption energy of new intermetallic systems and DFT is automatically performed on the most ideal candidates to verify the predictions. This process goes on iteratively in a closed feedback loop and 131 intermetallic surfaces across 54 alloys were identified as promising candidates for $CO_2$ reduction. Experimental validation[100] with Cu-Al catalysts yielded an unprecedented Faradaic efficiency of 80 % towards ethylene at a high current density of 400 mA cm$^{-2}$.

Due to these large range of variables that electrocatalysts may possess, it is consequently difficult to do data mining on the literature[101]. Many of these variables may not be fully characterized by research groups in their publications. Electrocatalyst structures are complex and difficult to completely characterize: efforts should be made to ensure that other factors that affect the electrocatalyst performance must be kept consistent. This is to avoid situations in which potentially promising compositions perform poorly as a result of non-ideal synthesis conditions or testing conditions. New approaches such as carbothermal shock synthesis[102,103] may be a promising avenue due to its propensity to generate uniformly sized and shaped alloy nanoparticles, regardless of composition.

X-ray absorption spectroscopy (XAS)[104] is a powerful technique, especially for *in-situ* measurements, and has been widely employed to gain crucial insight into the nature of active sites and changes in the electrocatalyst with time. The data analysis relies heavily on human experience and expertise, and as a result there has been interest in developing ML tools for interpreting XAS data.[35] Hung and co-workers developed improved RF models that can predict the Bader charge and nearest neighbour distances: crucial factors that influence catalytic properties of the material.[105] The EXAFS region of XAS spectra is known to contain information on bonding environments and coordination numbers. Frenkel and co-workers have shown that neural networks can be used to automatically interpret EXAFS data[36]. They decoded the structure of bimetallic nanoparticles using a neural network and experimental XAS data[106]. Raman and IR spectroscopy are also important tools for mechanistic understanding of electrocatalysis. Together with explainable AI, this could be used to discover descriptors hidden in spectra that will lead to new paradigms in electrocatalyst discovery and optimization.

*Fuel cell and electrolyser device management*

A fuel cell is an electrochemical device that can be used to convert chemical energy of a fuel (*e.g.*, hydrogen) into electrical energy; an electrolyser transforms electrical energy into chemical energy (*e.g.,* water splitting to generate hydrogen). ML has been employed to optimise/manage their performance, predict degradation/device lifetime as well as detect/diagnose faults. Djerdir and co-workers used a hybrid method consisting of extreme learning machine, genetic algorithm and wavelet analysis to predict degradation in proton exchange membrane fuel cells (PEMFCs)[107,108]. Kær and co-workers utilized electrochemical impedance measurements as input for an artificial neural network for fault detection and isolation in a high temperature PEMFC stack[109,110].

ML approaches can also be employed to diagnose faults in solid oxide fuel cell stacks such as fuel leakage and air leakage issues. Khan and co-workers used an artificial neural network to predict the performance of a solid oxide fuel cell under different operating conditions[111]. In addition, ML has been applied to optimise the performance of solid oxide electrolysers for $CO_2$/$H_2O$ reduction[112] as well as for chloralkali electrolysers[113].

## Incorporating XPIs

For each of the energy systems discussed above, the most commonly reported XPI is the acceleration factor (AF). This is in part because it is deterministic, but also because it is easy to calculate at the end of the development of a workflow. In most cases, we expect that authors report the AF only after completing development of the platform. Reporting the other XPIs will provide researchers with a better sense of both the time and the human resources that were required to develop the platform to the point where it was ready for publication. Moving forward, we hope that other researchers adopt the XPIs – or other similar metrics – to allow for fair and consistent comparison between the different methods and algorithms that are used to accelerate materials discovery.

## Integration of ML into smart power grids

A power grid is responsible for delivering electrical energy from producers (*e.g.*, power plants and solar farms) to consumers (*e.g.*, homes and offices). However, energy fluctuations from intermittent renewable energy generators can render the grid vulnerable[114]. ML algorithms can be used to optimize the automatic generation control (AGC) of power grids, which controls the power output of multiple generators in an energy system. When a relaxed Deep Learning (DL) model was used as a unified time scale controller for the AGC unit, the total operational cost was reduced by up to 80% compared to traditional heuristic control strategies[115]. A multi-agent reinforcement learning based smart generation control strategy was also found to improve the control performance by ~10% compared to other ML algorithms[43].

Accurate demand and load prediction can support decision-making operations in energy systems for proper load scheduling and power allocation. Multiple ML methods have been proposed to precisely predict the demand load: for example, long-short term memory was used to successfully and accurately predict hourly building load[42]. Short-term load forecasting of diverse customers (*e.g.*, retail businesses) using a deep neural network and cross-building energy demand forecasting via a deep belief network have also been demonstrated effectively[116,117].

Demand side management consists of a set of mechanisms that shape consumer electricity consumption by dynamically adjusting the price of electricity. These include reducing (*e.g.*, peak shaving), increasing (*e.g.*, load growth) and rescheduling (*e.g.*, load shifting) the energy demand, which allows for flexible balancing of renewable electricity generation and load[118]. A reinforcement learning-based algorithm resulted in significant cost reduction for both the service provider and customer[119]. A decentralized learning-based residential demand scheduling technique successfully shifted up to 35% of the energy demand to periods of high wind availability, resulting in significant power cost savings compared to the unscheduled energy demand scenario[120]. Load forecasting using a multi-agent approach integrates load prediction with reinforcement learning algorithms to shift energy usage (*e.g.*, electrical devices in a household) to optimize renewable energy usage[121]. This approach reduced peak usage by more than 30% and increased off-peak usage by 50%, reducing the cost and energy losses associated with energy storage.

## Areas of opportunity for machine learning and renewable energy

In this subsection, we will discuss areas in which ML has the opportunity to make significant further advances (**Fig. 5**). The subfields we discussed in the earlier sections are the branches of the energy materials field and they share similar materials-related challenges. We will also cover the grand challenges for ML application in smart grid and policy optimization.

*Representing materials with novel geometries*

A ML representation is effective when it captures the inherent properties of the system of interest (*e.g.*, physical symmetries) and can be utilized in downstream ancillary tasks:[122] transfer learning to new predictive tasks, building new knowledge *via* visualization or attribution, generating similar data distributions with generative models, *etc*.

For materials, the inputs are molecules or crystal structures whose physical properties are modelled by the Schrödinger equation (SE). Designing a general representation of materials that reflect these properties is an ongoing research problem. For molecular systems, several representations have been used successfully, including fingerprints,[123] SMILES,[20] SELFIES,[124] and graphs.[125–127] Representing crystalline materials has the added complexity of needing to incorporate periodicity in the representation. Methods like the smooth overlap of atomic positions,[128] Voronoi tessellation,[129,130] diffraction images,[131] multi-perspective fingerprints,[132] and graph-based algorithms[17,133] have been suggested, but typically lack the capability for structure reconstruction.

Complex structural systems found in energy materials present additional modelling challenges: 1) a large number of atoms (*e.g.*, reticular frameworks, polymers); 2) specific symmetries (*e.g.*, molecules with particular space group, reticular frameworks belong to certain topology); 3) atomic disordering, partial occupancy, or amorphous phases (*i.e.*, explore the enormous combinatorial space); 4) defects and dislocations (*e.g.*, interfaces, grain boundaries); 5) low dimensional materials (*e.g.*, nanoparticles). Reduction approximations alleviate the first issue (*e.g.*, RFcode for reticular framework representation),[9] while the remaining several problems warrant intensive future research efforts.

Self-supervised learning, which seeks to lever large amounts of synthetic labels and tasks to continue learning without experimental labels,[134] Multi-task learning,[135] in which multiple material properties can be modelled jointly to exploit correlation structure between properties, and Meta-learning,[136] which looks at strategies that allow models to perform better in new datasets or in out-of distribution data, all offer avenues to build better representations. On the modelling front, new advances in attention mechanisms,[137,138] graph neural networks[139] and equivariant neural networks[140] expand our range of tools to model interactions and expected symmetries.

*Data quality and more robust predictive models*

Predictive models are the first step when building a pipeline that seeks materials with desired properties. A key component for building these models is training data: often more data will translate into more performant models, which will translate into more correctly predicted new materials. DL models tend to scale more favourably with dataset size than traditional ML approaches (*e.g.*, random forests). Dataset quality is also essential. Experiments are usually conducted under diverse conditions with large variation in untracked variables. Additionally, public datasets are more likely to suffer from publication bias, as negative results are less likely to be published even though they are just as important as positive results when training statistical models.[11]

Addressing these issues require transparency and standardization of the experimental data being reported in literature. Text and natural language processing strategies could be then employed to extract data from the literature.[101] Data should be reported with the belief that it will eventually be consolidated in a database, such as the MatD$^3$ database.[141] Autonomous lab techniques will help address this issue.[23,25] Structured property databases such as Materials Project[11] and Harvard Clean Energy Project[12] can also provide a large amount of data. The different energy fields - energy storage, harvesting and conversion - should converge upon a standard and uniform way to report data. This should be continuously updated: as researchers continue to learn about the systems they are studying, conditions that were previously thought to be unimportant will become relevant.

New modelling approaches that work in low-data regimes are important, such as data-efficient models, dataset building strategies (active sampling)[142] and data-augmentation techniques.[143] Uncertainty quantification, data-efficiency, interpretability and regularization are important considerations that improve the robustness of ML models. These relate to the notion of generalizability: predictions should generalize to a new class of materials that is out of the distribution of the original dataset. The researcher can attempt to model how far away new data points are from its training set[144] or the variability in predicted labels with uncertainty quantification.[145] Neural networks are a flexible model class, and so often models can be under-specified,[146] and incorporating regularization, inductive biases or priors can boost the credibility of a model. Another effort to create trustable models is to enhance interpretability of ML algorithms by deriving feature relevance and score importance from them.[147] This could help identify potential chemically meaningful features and form a starting point for understanding latent factors that dominate material properties. These techniques can also identify the presence of model bias and overfitting as well as improve generalization and performance.[148–151]

*Synthesizing and evaluating new materials for stability*

The formation energy of a compound is used to estimate its stability and synthesizability. While negative values usually correspond to "stable" or "synthesizable", slightly positive formation energies below a limit[26,27] will lead to "metastable" phases with unclear synthesizability.[26,27] This is more apparent when investigating unexplored chemical spaces with undetermined equilibrium ground states; yet often the metastable phases exhibit superior properties (*e.g.*, photovoltaics,[154] ion conductors[155]). It is thus of interest to develop a method to evaluate the synthesizability of metastable phases. Instead of estimating the probability that a particular phase *can* be synthesized, one can instead evaluate its synthetic complexity using ML. In organic chemistry, synthesis complexity is evaluated based on the accessibility of their synthesis route[156] or precedent reaction knowledge.[157] Similar methodologies can be applied to the inorganic field with the ongoing design of automated synthesis planning algorithms for inorganic materials.[158,159].

Synthesis and evaluation of a new material alone does not ensure that material will make it to market: material stability is a crucial property that takes a necessarily long time to evaluate. Degradation is a generally complex process that occurs through the loss of active matter or growth of inactive phases (*e.g.*, rocksalt phases formed in layered Li-ion battery electrodes,[160] Pt particle agglomeration in fuel cell[161]) and/or propagation of defects (*e.g.*, cracks in cycled battery electrode[162]). Microscopies (*e.g.*, electron microscopy[163]) and simulations (*e.g.*, continuum mechanics modeling[164]) are commonly used to investigate growth and propagation dynamics *(i.e.*, phase boundary and defect surface movements versus time). However, these techniques are usually expensive and do not allow rapid degradation prediction. DL techniques such as convolutional neural networks (CNN) and recurrent neural networks (RNN) may be able to predict the phase boundary/defect pattern evolution under certain conditions after proper training.[165] Similar models can then be built to understand multiple degradation phenomena and aid the design of materials with improved cycle life.

*Optimizing the management of smart power grids*

A promising prospect of ML in smart grids is automating the decision-making processes that are associated with dynamic power supplies to distribute power most efficiently. Practical deployment of ML technologies into physical systems remains difficult due to data scarcity and a risk-averse mindset. The collection of and access to large amounts of diverse data is challenging due to high cost, long delay, and concerns over compliance and security.[166] For instance, to capture the variation of renewable resources due to peak/off-peak and seasonal attributes, long-term data collections are implemented for periods of 24 hours to several years.[167] While ML algorithms are ideally supposed to account for all uncertainties and unpredictable situations in energy systems, the risk-adverse mindset in the energy management industry still relies on human decision-making.[168]

An ML-based framework that involves a digital twin of the physical system can address these problems.[169,170] The digital twin represents the digitalized cyber models of the physical system and can be constructed from physical laws and/or ML models trained using data sampled from the physical system. This aims to accurately simulate the dynamics of the physical system, enabling relatively fast generation of large amounts of high-quality synthetic data at low cost. Notably, since ML model training and validation is performed on the digital twin, there is no risk to the actual physical system. Based on the prediction results, proper actions can be suggested and then implemented in the physical system to ensure stability and/or improve system operation.

*Bridging theory from materials to device to policy*

Finally, research generally is focused on one narrow aspect of a larger problem: we argue that energy research needs a more integrated approach.[171] Energy policy is the manner in which an entity (*e.g.*, the government) addresses its energy issues, including conversion, distribution, and utilization. ML has been used in the fields of energy economics finance. They have been employed for performance diagnostics (*e.g.*, oil well), forecasting energy generation (e.g., wind power) and consumption (*e.g.*, power load), predicting system lifespan (*e.g.*, battery cell life) and failure (*e.g.*, grid outrage).[172] They have also been used for energy policy analysis and evaluation (*e.g.*, estimating energy savings). A natural extension of ML models is to use them for policy optimizations,[173,174] a concept that has not yet seen widespread use. We posit that the best energy policies – including the deployment of the newly discovered materials – can be improved and augmented with ML and should be discussed in research reporting accelerated energy technology platforms.

## Conclusions

To summarize, ML has the potential to enable breakthroughs in the development and deployment of sustainable energy techniques. There have been remarkable achievements in many areas of energy technology, from materials design to device management to system deployment. ML is particularly well-suited to discovering new materials, and while the field is still nascent, there is conclusive evidence that ML is at least able to learn the same trends that human researchers have noticed over decades of research. Researchers in the field are expecting ML to bring up new materials which may revolutionize the energy industry. Meanwhile, the ML field itself is still seeing rapid development, with new methodologies being reported daily. It will take time to discover and adopt the methodologies that are specifically suited to solve problems in materials science. We believe that for ML to truly accelerate the deployment of sustainable energy, it should be deployed as a tool, similar to a synthesis procedure, characterization equipment or control apparatus. Researchers using ML to accelerate energy technology discovery should judge the success of the method primarily on the applied advance it enables. To this end, we have proposed the XPIs and a series of future directions in which we hope to see ML deployed.

# Figures

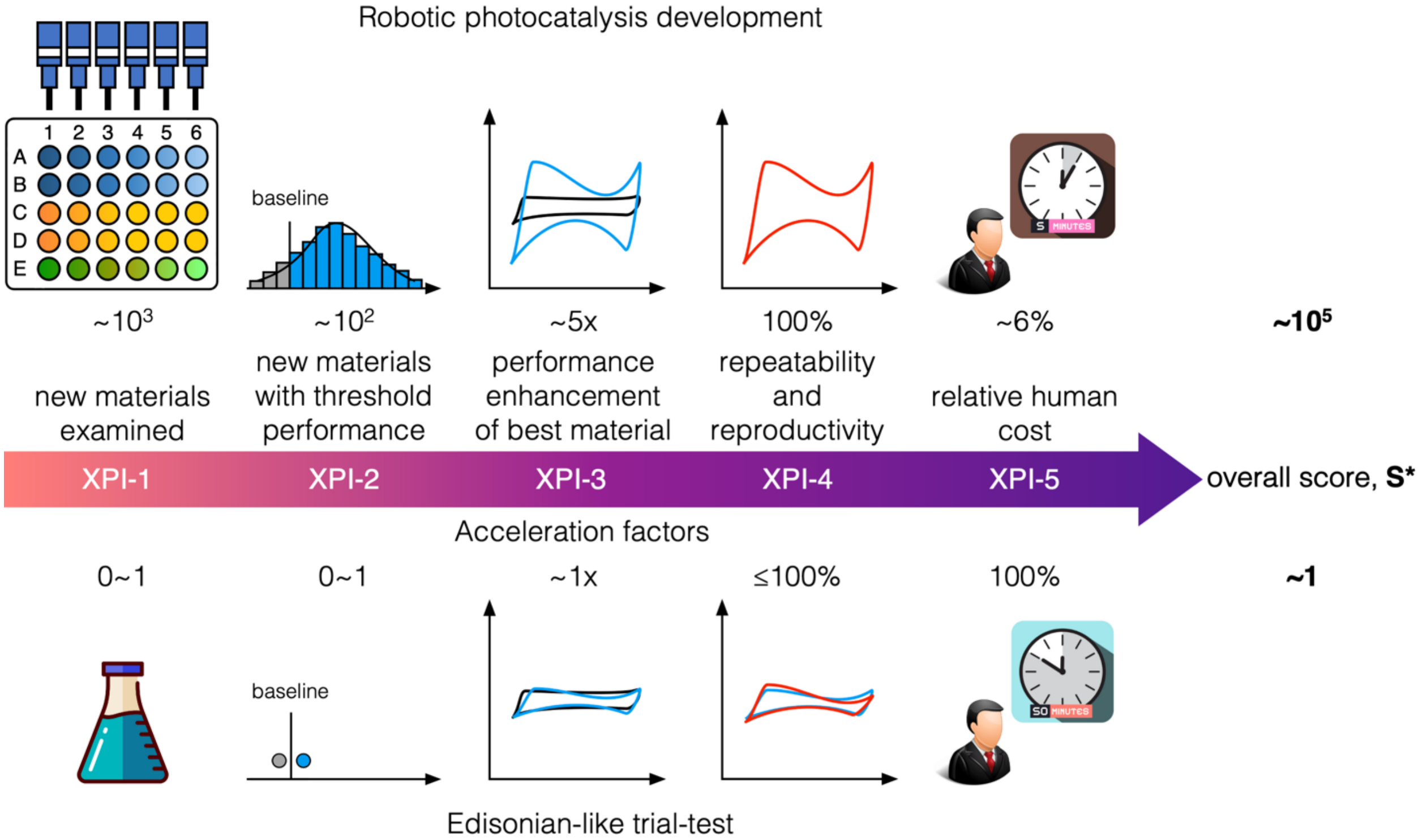

**Fig. 1 | Demonstration of the Acc(X)eleration Performance Indicators (XPIs).**

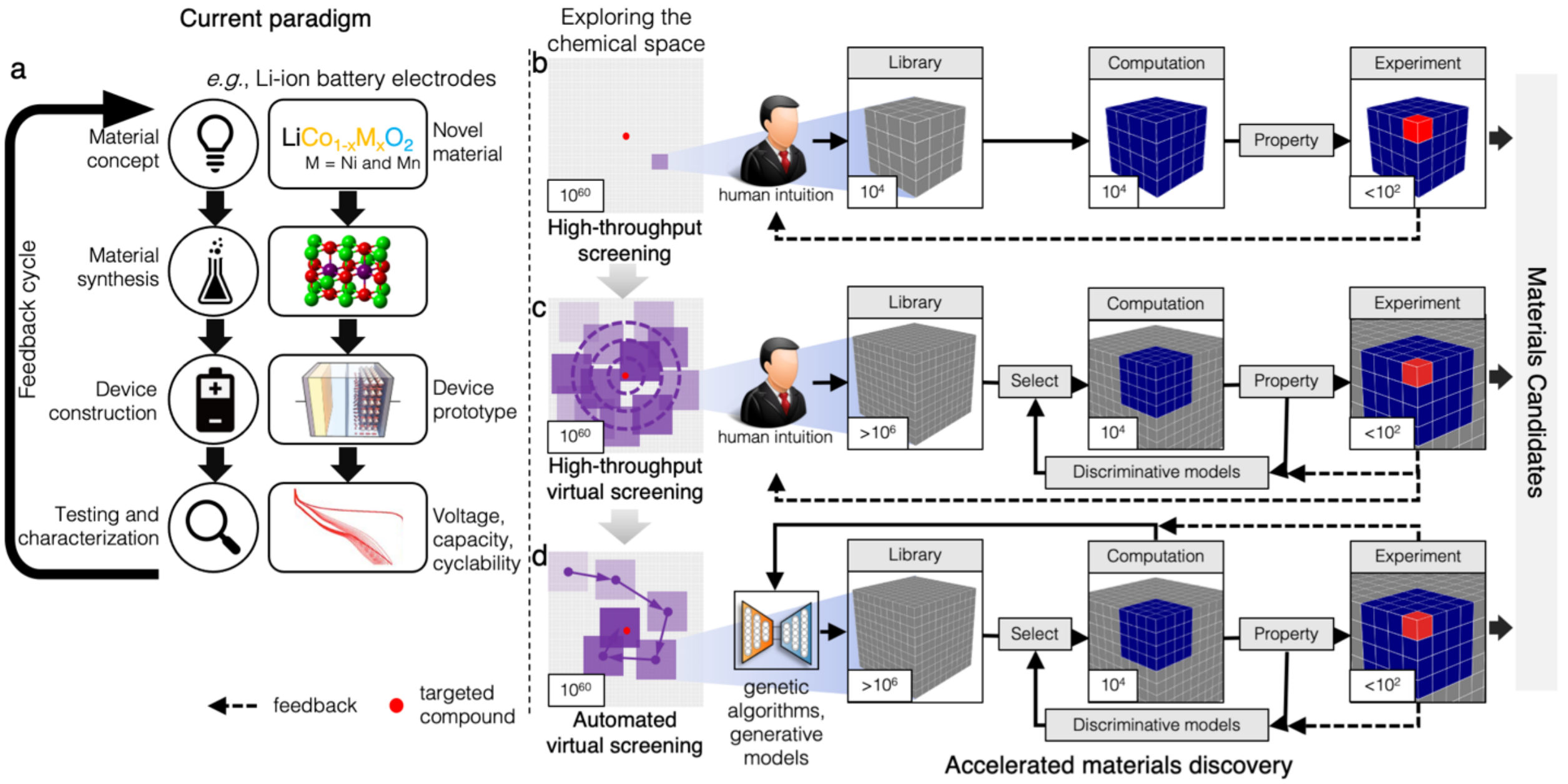

**Fig. 2 | Traditional and accelerated approaches to materials discovery.** (a) The traditional Edisonian-like approach, which involves experimental trial and error. (b) High-throughput screening approach involving a combination of theory and experiment. (c) ML driven approach whereby theoretical and experimental results are used to train a ML model for predicting structure and property relationships. (d) ML driven approach for property directed and automatic exploration of the chemical space using optimization ML (*e.g.*, genetic algorithms, generative models) that solve the 'inverse' design problem.

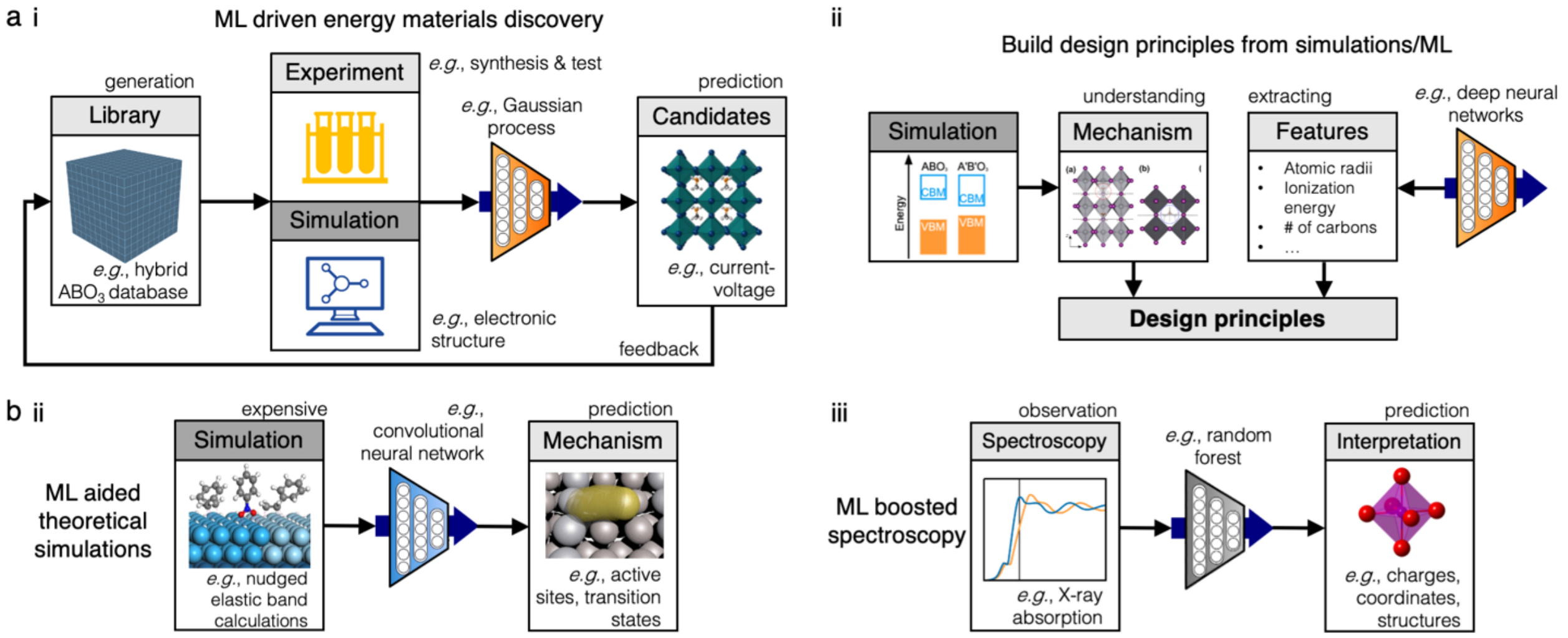

**Fig. 3 | Accelerating energy materials design and characterization.** a-i, illustration of the closed-loop energy materials design enabled by ML and simulations. a-ii, refinement of design principles from simulation and ML. b-i, ML aided theoretical simulations. b-ii, ML boosted spectroscopy.

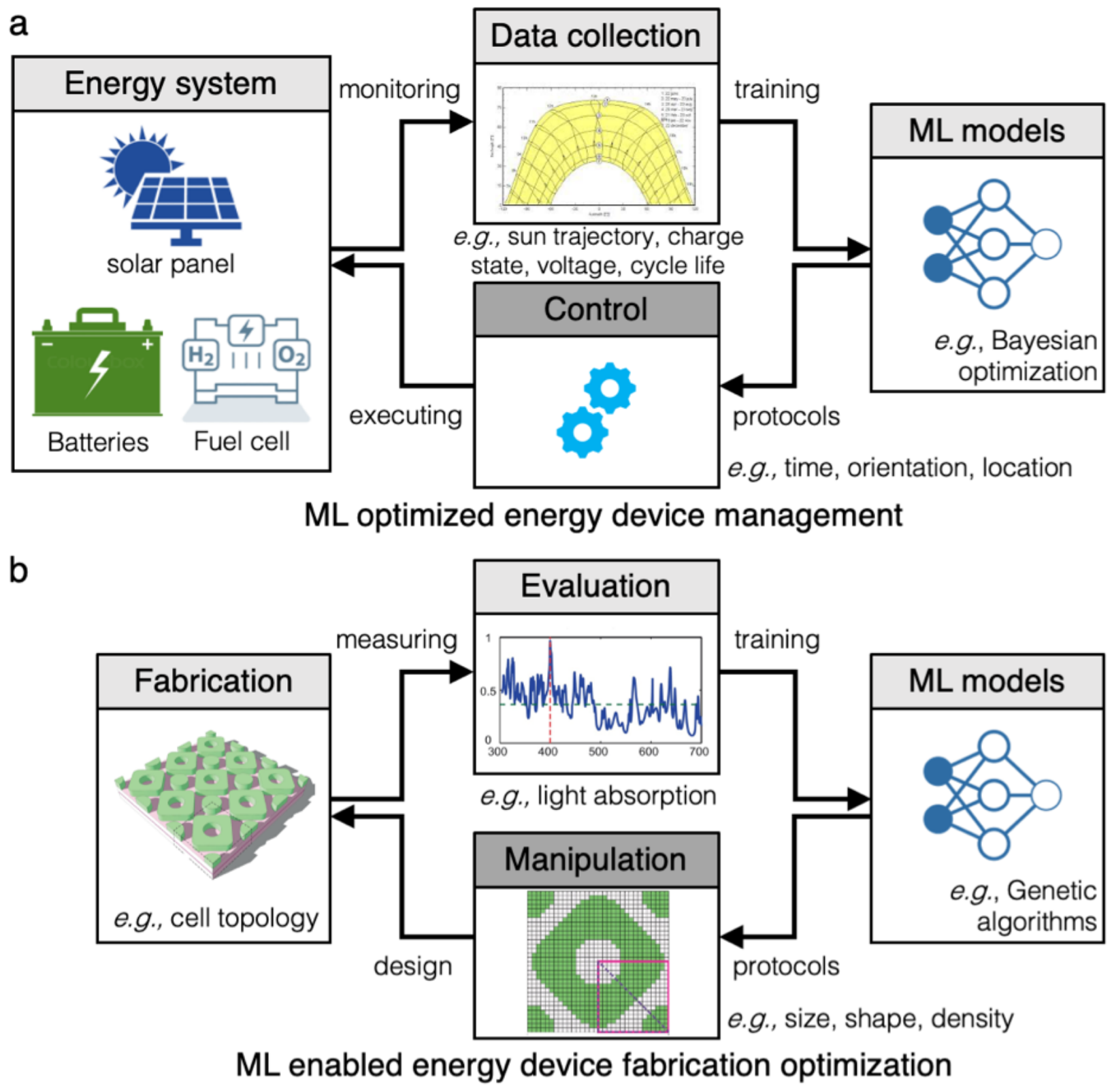

**Fig. 4 | ML optimized energy device management and fabrication**. a, illustration of the ML optimized energy device management. b, illustration of the ML enabled energy device design and fabrication.

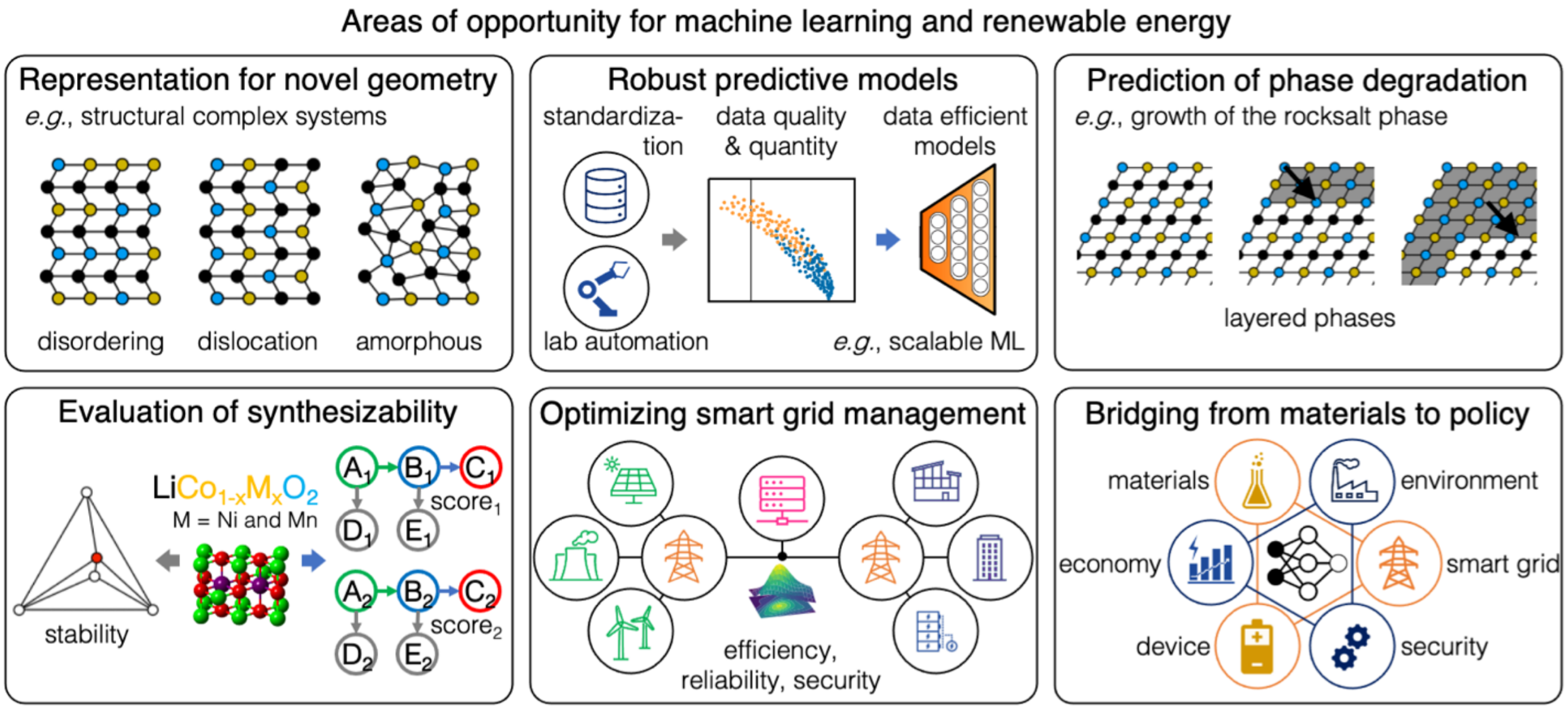

**Fig. 5 | Areas of opportunity for ML and renewable energy.** Opportunities upon extensive ML aided energy research.

**Tab. 1 | Grand challenges in materials research.** A list of some key grand challenges in energy research for photovoltaics, batteries and electrocatalysis, where ML could play a significant role in advancing.

| **Subfield** | **Grand challenges** | |
|---|---|---|
| | Materials[1,3] | Devices[1,3] |
| 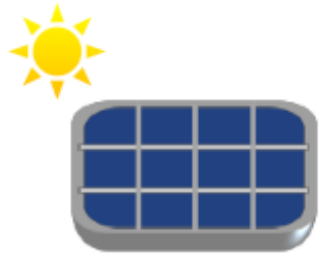 Photovoltaics | Discover non-toxic (Pd and Cd free) materials with good optoelectronic properties.<br>Identify and minimize materials defects in light absorber materials.<br>Design effective recombination layer materials for tandem solar cells.<br>Develop materials design strategy for long-term operational stability.<br>Develop (hole/electron) transport materials with high carrier mobility. | Optimize cell structure for maximized light absorption and minimized active materials usage.<br>Tune materials band gaps for optimal solar harvesting performance under complex operation conditions. |
| 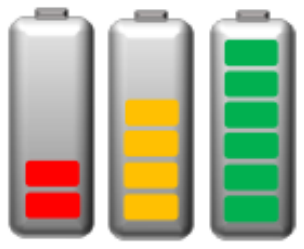 Batteries | Develop Earth abundant cathode materials (Co free) with high reversibility and charge capacity.[4]<br>Design electrolytes with wider electrochemical windows and high conductivity.<br>Identify electrolyte systems to boost battery performance and lifetime.<br>Discover new molecules for redox flow batteries with suitable voltage. | Understand correlation between defects growth in battery materials and overall degradation process of battery component.<br>Tune operando (dis)charging protocol for minimized capacity loss, (dis)charging rate and optimal battery life under diversified conditions.[7] |
| 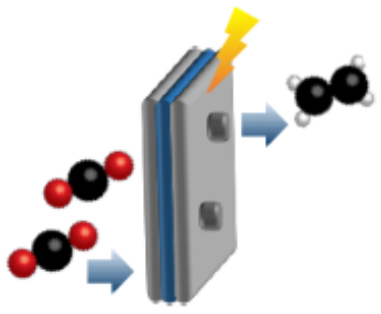 Electrocatalysis | Design materials with optimal adsorption energy for maximized catalytic activity.<br>Identify and study active sites on catalytic materials.<br>Engineer catalytic materials for extended durability.<br>Identify a fuller set of materials descriptors that relate to catalytic activity.<br>Engineer multiple catalytic functionality into the same material. | Design multiscale electrode structure for optimized catalytic activity.<br>Correlate atomistic contamination and growth of catalyst particles to electrode degradation process.<br>Tune operando (dis)charging protocol for minimized capacity loss and optimal cell life. |

**Tab. 2 | Summary of advances of applying ML approach to energy harvesting, storage, and conversion.**

| Research area | ML approach | Summary of main research outcome | Ref. |
|---|---|---|---|
| Photovoltaics | Bayesian optimisation | By sampling just 1.8% of the compositional space, perovskites identified show >17-fold stability improvement over original MAPbI3 without compromising conversion efficiency. | 29 |
| | Random forest classifier | Approach reliability is further verified by screening 10 newly designed donor materials with good consistency between model predictions and experimental outcomes. | 175 |
| | ML regression algorithms | Six lead-free hybrid perovskites with suitable bandgaps for solar cells and room temperature thermal stability were successfully screened out from 5158 candidates. | 54 |
| | Gaussian process regression | The ML model was able to make bandgap predictions of elpasolite compounds with similar accuracy to that of high cost computational calculations. | 52 |
| | Kernel ridge regression (KRR) | Based on a set of 1.2 million features, two of them were identified as the most important factors that influence bandgap | 53 |
| | Random forest regressor | An acceleration factor of 700 compared to brute-force method was achieved with an experimentally validated new perovskite. | 55 |
| | DNN based classifier | The ML model can classify compounds 10 times faster than human analysis with 90% accuracy with four lead-free layered perovskites realized experimentally. | 176 |
| | CNN based classifier and random forest regressor | CNN based crystal recognition enabled autonomous characterization of the outcomes of the robotic experiments. The regressor predict the optimal conditions for the synthesis of a new perovskite single crystal. | 177 |
| | Bayesian optimisation | Using an automated experimentation platform with a Bayesian optimization, a 4D parameter space of organic photovoltaics blends is mapped and optimized for photostability | 178 |
| | Random forest regressor | Essential features were identified and used for the screening of the most promising capping layer, which then leads to stability increase of state-of-the-art cell by multiple times. | 179 |
| | Random forest regressor | Major patterns regarding materials selection/device structure were captured which can be used to predict solar cell efficiencies. | 180 |
| | Genetic algorithm | Experimental samples processed under conditions suggested by the model show significant improvements in performance. | 181 |
| Batteries | DNNs, KRR, and support vector machine | The model enabled a reduction in the amount of density functional theory (DFT) calculations to explore the chemical space, from which up to 5000 candidate materials for Na-ion and K-ion electrodes were identified. | 73 |
| | Artificial neural network | The model demonstrates a capability for accurately estimating the redox potentials with the contribution analysis confirming electron affinity has the highest contribution. | 182 |
| | Gaussian KRR and gradient boosting regression (GBR) | The method predicted the redox potentials well and the redox potentials can be explained by a small number of features, improving the interpretability of the results. | 183 |
| | Logistic regression | The screening reduces the list of candidate materials from 12831 down to 21 structures that show promise as electrolytes, few of which have been examined experimentally. | 66 |
| | Linear regression and support vector machine | The method transferred physical insights onto more generic descriptors, allowing the screening of billions of unknown compositions for Li-ion conductivity. | 184 |
| | Logistic regression | The ML-guided search is 2.7 times more likely to identify fast Li ion conductors, with at least a 44 times improvement of room temperature Li ion conductivity. | 185 |
| | Hierarchical and spectral clustering | Ab-initio molecular dynamics (AIMD) simulations were then used to validate the clustering and identifying top candidates, with 16 new Li-ion conductors discovered. | 68 |
| | Artificial neural network | Predicted electrode specific resistances were found to agree well with the simulated values. | 186 |
| | Crystal graph CNN, KRR and GBR | ML model was used to screen over 12000 inorganic solids for their use as solid electrolytes with four of them can be used to suppress Li dendrite growth. | 187 |
| | Model-free reinforcement learning | The method can explore the trade-off in the power-performance design space and converge it to a better power management policy. Experimental results with this technique exhibit a remarkable power reduction comparing to the existing expert-based power management. | 188 |
| | Bagged decision tree | The model leaded to a policy significantly outperforms the leading algorithms and the policy is capable of improving and adapting as new data is being collected over time. | 189 |
| Electrocatalysis | Random forest regressor and extra trees regressor | The framework was able to identify 131 intermetallic surfaces across 54 alloys as promising candidates for $CO_2$ reduction. Specifically, a Cu-Al alloy catalyst was identified and experimentally verified to selectively convert $CO_2$ to ethylene with record performance. | 99,100 |
| | Neural networks | The model reduced the number of intermediate ab initio calculations needed to locate saddle points during a nudged elastic band simulation. | 38 |
| | Gaussian process regressor | The model predicted the most important reaction step to be calculated with computationally demanding electronic structure theory and with this method the most likely reaction mechanism for Rd (111) was identified. | 37 |
| | Neural networks | This neural network was then able to screen for active sites across a random, disordered nanoparticle surface. The most likely active sites for $CO_2$ conversion were identified for Au and Cu nanoparticle systems. | 95,96 |

## Acknowledgments

**Z.Y.** and **A.A.-G.** were supported as part of the Nanoporous Materials Genome Center by the U.S. Department of Energy, Office of Science, Office of Basic Energy Sciences under award number DE-FG02-17ER16362 and the US Department of Energy, Office of Science - Chicago under Award Number DE-SC0019300. **A.J.** was financially supported by Huawei Technologies Canada Co., Ltd. and the Natural Sciences and Engineering Research Council (NSERC). **L.M.M.-M.** thanks the support of the Defense Advanced Research Projects Agency under the Accelerated Molecular Discovery Program under Cooperative Agreement No. HR00111920027 dated August 1, 2019. **Y. W.** acknowledges the funding support from the Singapore National Research Foundation under its Green Buildings Innovation Cluster (GBIC Award No. NRF2015ENC-GBICRD001-012) administrated by the Building and Construction Authority, its Green Data Centre Research (GDCR Award No. NRF2015ENC-GDCR01001-003) administrated by the Info-communications Media Development Authority, and its Energy Programme (EP Award No. NRF2017EWT-EP003-023) administrated by the Energy Market Authority of Singapore. **A.A.-G.** is a Canadian Institute for Advanced Research (CIFAR) Lebovic Fellow. **E.H.S.** acknowledges funding by the Ontario Ministry of Colleges and Universities (Grant ORF-RE08-034), Natural Sciences and Engineering Research Council (NSERC) of Canada (Grant RGPIN-2017-06477), Canadian Institute for Advanced Research (CIFAR) (Grant FS20-154 APPT.2378) and University of Toronto Connaught Fund (Grant GC 2012-13). **Z.W.S.** acknowledges funding by the Singapore National Research Foundation (NRF-NRFF2017-04).

## Author contributions:

Z.Y., Y.L., A.J. contributed equally to this work. All authors contributed to the writing and editing of the manuscript.

## Competing financial interests

The authors declare no competing interests.